\documentclass[a4paper,11pt]{article}%
\usepackage{amsfonts}
\usepackage{amssymb}
\usepackage{amsmath}
\usepackage{verbatim}
\usepackage{url,natbib}

\makeindex

\makeatletter

\renewcommand*{\@fnsymbol}[1]{\ensuremath{\ifcase#1\or \star\or a\or b\or
   \mathsection\or \mathparagraph\or \|\or \star\star\or \dagger\dagger
   \or \ddagger\ddagger \else\@ctrerr\fi}}

\makeatother

\begin{document}
\title{Deep Learning in (and of) Agent-Based Models:\\A Prospectus\thanks{This paper has benefited from discussions with Spyros Kousides, Nan Su, Herbert Dawid and Blake LeBaron. Any remaining errors or omissions are the sole responsibility of the author. Financial support from the Horizon 2020 ISIGrowth Project (Innovation-fuelled, Sustainable, Inclusive Growth), under grant no. 649186, is gratefully acknowledged.}}

\author{Sander van der Hoog\thanks{Computational Economics Group, Department of Business Administration and Economics, Chair for Economic Theory and Computation Economics, Bielefeld University, Universit\"atsstrasse 25, 33615 Bielefeld, Germany. Email: svdhoog@wiwi.uni-bielefeld.de.}}

\date{\today}

\maketitle

\begin{abstract}
A very timely issue for economic agent-based models (ABMs) is their empirical estimation. This paper describes a line of research that could resolve the issue by using machine learning techniques, using multi-layer artificial neural networks (ANNs), or so called Deep Nets. The seminal contribution by Hinton et al. (2006) introduced a fast and efficient training algorithm called Deep Learning, and there have been major breakthroughs in machine learning ever since. Economics has not yet benefited from these developments, and therefore we believe that now is the right time to apply Deep Learning and multi-layered neural networks to agent-based models in economics.
\end{abstract}



\clearpage
\section{Introduction}

Agent-Based Models (ABMs) are becoming a powerful new paradigm for describing complex socio-economic systems. A very timely issue for such models is their empirical estimation. The research programme described in this paper will use machine learning techniques to approach the problem, using multi-layer artificial neural networks (ANNs), such as Deep Belief Networks and Restricted Boltzmann Machines. The seminal contribution by Hinton et al. (2006) introduced a fast and efficient training algorithm called Deep Learning, and there have been major breakthroughs in machine learning ever since. Economics has not yet benefited from these developments, and therefore we believe that now is the right time to apply Deep Learning and multi-layer neural nets to agent-based models in economics.

Economic Science is undergoing its own form of "climate change" in economic theory as new subfields such as behavioural, experimental, computational, and complexity economics are gaining in support. Complexity economics brings in new tools and techniques that were originally developed in physics and computer science, such as the theory of networks and new statistical techniques for the study of many-body dynamics.

The agenda of this paper is to briefly sketch the current state-of-the-art in Artificial Intelligence (AI) and Machine Learning (ML), and apply them to economic decision-making problems. We outline a research programme that encompasses co-evolutionary learning, learning from experience, and artificial neural networks (ANN), and connect these to agent-based modelling in economics and policy-making.

We begin by tracing back the common heritage of complexity economics and evolutionary economics. A rich body of work by evolutionary economists deals with decision-making by real economic agents, in an attempt to capture their routines by automatizing their decision-making processes in computer algorithms. This starts with Herbert Simon's "A Behavioral Model of Rational Choice" (\citealp{Simon:1955}), it continues with Cyert and March's "A Behavioral Theory of the Firm" (\citealp{CyertMarch:1963}), and culminates in Simon's seminal work on "The Sciences of the Artificial" (\citealp{Simon:1969}). 

Around the same time, a computer science conference on "The Mechanization of Thought Processes" was held at the National Physical Laboratory (NPL) in 1958 (\citealp{NPL:1958}). The conference proceedings contain many of today's hot topics in Machine Learning: automatic pattern recognition, automatic speech recognition, and automatic language translation. At the time, the developments in AI, machine learning and theories of human decision-making were strongly intertwined. 

After describing this common heritage, we take stock of the current state-of-the-art in Machine Learning and extrapolate into the not too distant future.

Firstly, we propose to emulate artificial agent behaviour by so called surrogate modelling, which could be thought of as a Doppelg\"anger approach, in which one agent is observing another agent's behavioural pattern and their performance. It then tries to imitate that agent, and eventually replace it in the simulation. In addition, the concept of an ANN-Policy-Agent is introduced who learns from observations of successful policy actions through reinforcement learning mechanisms.

Secondly, we propose to use ANNs as computational emulators of entire ABMs. The ANN functions as a computational approximation of the non-linear, multivariate time series generated by the ABM. It is a meta-modelling approach using statistical machine learning techniques. There are various advantages to having such an emulator. It allows for a computationally tractable solution to the issue of parameter sensitivity analysis, robustness analysis, and could also be used for empirical validation and estimation. This is particularly appealing for large-scale ABMs that are computationally costly to simulate.

The goal is to develop new computational methods to improve the applicability of macroeconomic ABMs to economic policy analysis. When successful, we would have drastically reduced the complexity and computational load of simulating ABMs, and come up with new methods to model economic agents’ behaviour. Linking the time series forecasting capabilities of the Deep Learning algorithm to ABMs also allows us to envision the possibility of docking experiments between different ABMs: the time series output from one ABM can be fed into the Deep Learning algorithm, resulting in an artificial neural network. This artificial
neural network can then be used as an agent inside another, larger-scale ABM. This notion leads to a hierarchical modelling scheme, in which ABMs of ABMs would become feasible.

Each agent in the larger ABM can have an internal ”mental model” of the world it inhabits, and those mental models can differ to any degree. On the longer term, this approach would allow the inclusion of computational cognitive models into economic ABMs, allowing the agents to be fully aware of their environment, and to consider the social embedding of their interactions.

\subsection{Related literature}
In many cases, we do not know the correct equations of the economic model, and we might only know the behaviour of the artificial economic agents approximately through observations of the empirical behaviour of their real-world counterparts (e.g., through direct observation of market participants, or through laboratory experiments). Therefore, we do not have access to the mathematical description of the economic system, and have to resort to computational modelling. 
Once we have constructed a computational model that satisfies certain requirements (e.g., stock- flow consistency of accounting relationships or dynamic completeness of behavioural repertoires) we usually find that the model is realistic enough to reproduce several empirical stylized facts of macrovariables, such as GDP growth rates, inflation rates, and unemployment rates, but all too often it is computationally heavy.

There are currently several research efforts under way to construct agent-based macroeconomic models (\citealp{Dawid:2014, Dosi:2014, Grazzini:2013c}). These models aim to compete with standard Dynamic Stochastic General Equilibrium (DSGE) models that are currently in use by ECOFIN and most Central Banks around the world. Using such models for policy analysis requires that they are calibrated and estimated on empirical data. For this, we need new methods and techniques to estimate such policy-oriented ABMs.

Current large-scale agent-based simulation models (e.g., \citealp{Dawid:2014}) require large computing systems, such as multi-processor servers, HPCs, or grids of GPUs, in order to run sufficiently many simulations. This not only involves running large numbers of simulations for producing results for publications, but also to perform rigorous robustness testing, parameter sensitivity analyses, and general verification and validation (V\&V) procedures to ensure the correctness and validity of the computer simulations (cf. \citealp{Sargent:2011, Kleijnen:1995}).

The issue of computational intractability is ubiquitous. It has been around for a long time in physics and climate science, where research using many-particle systems and large-scale climate models is constantly pushing the frontier of what is feasible from a computational point of view.

In this paper we describe how to tackle the problem by taking advantage of machine learning techniques, in particular recent developments in artificial neural networks, such as Deep Learning, Deep Belief Networks, Recursive Networks and Restricted Boltzmann Machines.

The scientific relevance and innovativeness of this line of research is that it tries to solve the generic problem of computational tractability of computer simulation models (and more specifically, of agent-based economic models), not by resorting to technological solutions (e.g., parallel computing, or GPU grids), but by using machine learning algorithms in order to reduce the computer simulation to a lighter form, by emulating the models using artificial neural networks, and by then adopting that simulation model to obtain results.

In order for agent-based models to be useful for economic policy analysis so called "what-if scenarios" are used to test counter-factuals in would-be worlds. It is therefore necessary to use models with a sufficiently high resolution in terms of the behavioural and institutional details. The target models that we consider in this paper are large-scale agent-based models where 'large-scale' means on the order of millions of agents. High-resolution may refer to the resolution of time-scales, geographical scales, decision-making scales (number of options to consider), or other dimensionality of the agents' characteristics. 

The advantage of such large-scale, high-resolution, high-fidelity agent-based models is that they can be used as virtual laboratories, or as laboratory "in silico" (\citealp{TesfatsionJudd:2006}). The model can be used for testing various economic policies (\citealp{DawidFagiolo:2008, DawidNeugart:2011, FagioloRoventini:2012a, FagioloRoventini:2012b}), that may not be feasible to test in the real world (e.g., due to ethical objections). Examples include: What happens when the biggest banks go bankrupt? Or: What happens when a Euro member leaves the Euro? Obviously, these are not things we want to simply test in the real world, considering the detrimental social consequences and ethical objections.
The disadvantage is that such large-scale ABMs are quite heavy from a computational perspective. It is easy to generate overwhelming amounts of data, and reach the boundaries of what is commonly accepted to be computationally tractable, in terms of simulation time, number of processors used, and data storage requirements. 

If we want to apply such models to perform policy analyses, we have to test the robustness of the model, i.e., to test whether the empirical stylized facts are still reproduced for many parameter settings. This involves performing a global parameter sensitivity analysis and a robustness analysis against small changes in the economic mechanisms, or with respect to changes in the individual behavioural repertoires of the agents. This usually requires a large number of simulations (on the order of thousands), in order to obtain a large enough sampling of the phase space, and to be able to ascertain whether the model is sensitive, stable, robust, or fragile. 

In the social sciences where computer simulation models are being actively pursued (e.g., economics, sociology, econophysics) there are many discussions surrounding the empirical estimation and validation of these types of models (e.g., \citealp{Werker:2004, Brenner:2006, Fagiolo:2007, Grazzini:2012, Grazzini:2013a, Grazzini:2013b, YildizogluSalle:2012, Barde:2015, Lamperti:2015}). However, until now, no clear consensus has appeared how to resolve the  the empirical validation problem. In econometric applications, some advances have been made on the estimation of ABMs. Noteworthy are two approaches, one using non-parametric bootstrap methods (\citealp{Boswijk:2007}) and the other using estimation of a master equation derived from the Focker-Planck equation (\citealp{Alfarano:2005, Aoki:2007, DiGuilmi:2008}).

Currently, multiple projects are under way to construct agent-based macroeconomic models: the Eurace@Unibi model (\citealp{Dawid:2014}), the Crisis Project (\citealp{Grazzini:2013c}), and the "Keynes meeting Schumpeter" models (K+S models, \citealp{Dosi:2010,Dosi:2013,Dosi:2014}). These models consider it a feature, not a vice, to model the agents and their behavioural repertoires in great detail, by taking care that all the behavioural assumptions are supported by empirical evidence.

\section{Applying machine learning methods to economic problems}
A primary motivation for applying machine learning techniques to economic decision making problems is the work by Herbert Simon on bounded rationality and satisficing in \textit{"Administrative Behavior"} and \textit{"Sciences of the Artificial"} (\citealp{Simon:1947,Simon:1955,Simon:1959,Simon:1969}). Simon being also one of the founders of modern-day Artificial Intelligence (AI), it seems only appropriate that in applying artificial neural networks to economic problems, we rely on various aspects of Simon's path-breaking work.

The first aspect we adopt is goal-oriented, adaptive behaviour. In a perfect world agents are not required to spent time on planning and learning. They already have all the relevant information available, and are able to compute with full accuracy the outcome of their actions. However, a substantial amount of time of real decision makers is being spent on planning and learning about new information. Time constraints are important for making decisions, hence satisficing with threshold aspiration levels rather then optimizing would be the preferred methodology.

Open, complex systems  make it essential to behave in a flexible, adaptive manner, rather than using rigid, predetermined rules that prescribe an exact course of action for every contingency. This naturally leads to the use of heuristics, routines, and rules of thumb. Satisfying aspiration levels instead of optimizing appears to be more appropriate as a model of man, as in the adage 'Better to be approximately right, rather than exactly wrong.' Such an approach would lead to decision makers who realize they are fallible, and in order to achieve their goals they must do the best they can given the circumstances. They would aim for robust decision making routines, rather than precise prescriptions.

Such considerations point into the direction that human decision makers are not always able to make perfect decisions, due to various limitations in their decision making capabilities:
(i) Imperfect information gathering, or incomplete observation of outcomes.
(ii) Limitations in storage capacity or faulty interpretation of those observations (imperfect recall, bad documentation of results).
(iii) Limits in processing abilities.
(iv) Imperfections in foreseeing the exact consequences of their actions. Even when acting in perfect isolation or when they act in the belief that they have precise control over their actions, unintended consequences of deliberate, decisive human action may result from a noisy environment. All such imperfections in gathering, storing and processing of information and in foreseeing events are a fact of life for the human decision maker.

A second motivation for applying machine learning techniques to economic problems is the seminal book \textit{"A Behavioral Theory of the Firm"} by \citet{CyertMarch:1963}. This book describes many operating procedures related to real firm decision making. Besides an emphasis on organizational processes and decision making routines, a further aim of the theory was to link empirical data to the models by considering the results of case studies of real firms.

A clear assessment of the impact of \textit{A Behavioral Theory of the Firm} and its methodological stance was given by \citet[p.339]{ArgoteGreve:2007}:
\begin{quote}
"The general methodological point was that theory should model organizational processes, and should be generated through systematic observation of processes in actual organizations. One component of this point is that organizational theory should not oversimplify. Although parsimony is needed in theory building, parsimony that throws out basic insights -- like replacing a process model with a maximization assumption -- can be harmful."
\end{quote}

In the context of agent-based economic models, this idea has been developed further into the Management Science Approach (\citealp{DawidHarting:2011,Dawid:2014}). In this approach the economic agents are assumed to use decision making routines that are empirically-grounded in the Management Science literature. The underlying assumption is that managers of a firm apply the methods and techniques that they have been taught whilst doing their MBA at Management School. This method can for example be applied to model the pricing behaviour of firms, the inventory management problem, the interest rate setting for loans by bank managers, or the hiring and firing practices of a Human Resource Management department.

\smallskip
In our approach, we use a combination of the artificial intelligence methods proposed by Simon (learning appropriate heuristics in order to satisfy certain goals), and the empirically-grounded behavioural rules as proposed by Cyert and March (actual organizational processes).

\smallskip
Another exciting field of research is to include a rich cognitive structure into the agents' behavioral repertoires. The decision making routines, although adaptive, are often still too rigid from a cognitive science point of view. A lack of meta-rules to update the behavioral rules is often seen as a serious drawback of these models, especially when it comes to addressing the Lucas Critique which states that economic policy has to take into account the adaptive behavioral response by the agents that are subject to the policy.

This perceived lack of cognitive modelling in the behavioral routines of economic agents can be alleviated if we would allow each agent in the ABM to have an internal "mental model" of the world it inhabits, and those mental models can differ to any degree. On the longer term, this approach would allow the inclusion of computational cognitive models into economic agent-based models, allowing the agents to be fully aware of their environment, and possibly also to consider the social embedding of their interactions.

\subsection{Machine learning for time series forecasting in Economics}
Applications of ANNs to time series forecasting problems in economics include: financial market forecasting (\citealp{Trippi:1993, Azoff:1994, Refenes:1995, Gately:1996}), foreign exchange rates (\citealp{Weigend:1992, Refenes:1993, Kuan:1995}), load demand forecasts on electricity markets (\citealp{Bacha:1992, Srinivasan:1994}), commodity prices (\citealp{Kohzadi:1996}), and macroeconomic indices (\citealp{Maasoumi:1994})
A review of applications of ANNs in the field of Management Science and Operations Research is given by \citet{WilsonSharda:1992} and \citet{Sharda:1994}.
The M-competition (\citealp{Makridakis:1982}) provides a widely cited data base for comparing the forecasting performance of ANNs in comparison to traditional statistical methods. The data for the  M-competition are mostly from business, economics, and finance, see \citet{Kang:1991, Sharda:1994, Tang:1993} for examples.
Another comparison is provided by the Santa Fe forecasting competition (\citealp{Weigend:1993}) which includes very long time series coming from various fields.

\subsection{The usefulness of ANNs to study complexity science}
According to \citet{Gorr:1994}, ANNs are very appropriate in the following situations: (i) large data sets; (ii) problems with nonlinear structure; (iii) multivariate time series forecasting problems. Important issues that can be addressed include:
\begin{enumerate}
\item[(1)] How do ANNs model the autocorrelated time series data and produce better results than traditional linear and non-linear statistical methods?

According to \citet{Bengio:2013}, sequential statistical data (a.k.a. time series) suffer from the "difficulty of learning long-term dependencies". If the past is coded linearly
(regardless of how many observations in the past) then the effect of the past of the \emph{previous} step is diminishing 
exponentially. If the past is modelled non-linearly, then the non-linearity is "composed many times", leading to a highly non-linear relationship of past to present.
According to the paper cited, recurrent neural networks (RNN) are better in modelling such relationships. However, RNNs suffer from the problem of "diminishing gradients" when using 
back-propagation for training the weights with Stochastic Gradient Descent. In such cases Hessian-Free (HF) optimization methods or Momentum Methods such as Nesterov's Accelerated Gradient (NAG) seem more promising (see Section \ref{Section: Research themes}, Theme 3 for more details).

The paper suggest a number of optimizations for RNNs. We believe one of the most relevant for our problem is that of filtered, low-pass filter inputs. These are nodes with self-weights close to 1, (similar to 
exponential filters) that allow non-linearities to persist longer and not disappear at the next step. This is coupled with non-linearities modeled e.g. as $out = max(0, in)$ rather than a sigmoid or 
tanh function. There is justification for the approaches and some promising results (although this is by 
no means a solved problem) in that the output of the error function is "rough" and requires some form of control for local cliffs that lead to local minima. All of the methods proposed in the literature are gradient-tracking in one way or the other, and are conservative about sudden changes. Hessian-free optimization (\citealp{Martens:2011}) and the PhD-thesis by \citet{Sutskever:2013} show the applicability of such methods in the multivariate time series domain.

\item[(2)] Given a specific forecasting problem, how do we systematically build an appropriate network that is best suited for the problem?

We follow current best-practices as outlined above. We can start from the simplest RNN representation,
and try state-of-the-art approaches.
The design of good initializations of the networks is a good point of entry. If we have domain knowledge about the units that
operate in the system and their qualities, we can estimate the relative size of each input node and the
long term effect that actions should have. We then use state-of-the-art parameter estimation 
techniques, as described in \citet{Bengio:2013}, for example, in order to fix the weights on the input nodes.

\item[(3)] What is the best training method or algorithm for forecasting problems, particularly time series forecasting problems?

This is discussed extensively in \citet{Sutskever:2013}, noting that optimized Stochastic Gradient Descent (SGD) may be adequate, if one 
considers proper initialization of the network. Momentum methods are another option. As described above, we could start with the 
simplest method first (SGD), or consider the best practice for problems similar to ours. We should keep in mind, however, that we may
have various problems that are different in structure. The doppelganger ANN described in Theme 1 (micro-emulation) is not an RNN, but is
rather an actuator based on the inputs. It \emph{can} have memory of its own actions, but it is still distinctively different from an RNN that models a time series. Hence, we should find different best practices for each of our sub-tasks. In the theme descriptions in Section \ref{Section: Research themes} we make initial propositions for each of the theme descriptions.

\item[(4)] How should we go about designing the sampling scheme, and the pre- and post-processing of the data? What are the effects of these factors on the predictive performance of ANNs?

One of the advantages of ANNs is that they alleviate the need for \emph{feature engineering} which
is the art and science of traditional machine learning. Instead, any real number goes through a squashing function (logistic or tanh), resulting in a number between 0 and 1 (or -1 and 1). In case of categorical values, one can have a 'softmax layer', that assigns a probability distribution over the states. Alternatively, one can have "ON/OFF" nodes with binary values. The fine art then becomes how to design the structure of the network itself: how many layers, and how many nodes per layer.

\end{enumerate}
All these questions are addressed in more detail below.

\section{Applying Deep Learning algorithms to economic ABMs}
\label{Section: Research themes}

In order to use macroeconomic agent-based models for policy, we need to reduce the complexity of the ABM simulation to a less complex, more computationally tractable system. In other words, surrogate models or meta-modelling approaches need to be developed, that allow us to approximate or 'emulate' the multi-dimensional nonlinear dynamics of the original system. The entire process of finding such an approximate 'emulator' for an ABM consists of a four-step procedure.

First, we construct an ABM and generate synthetic time series data. Then, a multi-layered, deep neural network is designed and trained on the synthetic data.
Third, the trained neural network should be empirically validated using real-world data. And fourth, we apply the trained, empirically validated deep neural network to economic policy analysis.

According to these four steps, the question how to construct efficient emulators of ABMs could be structured along four broad research themes:
\begin{list}{}{}
\item[Theme 1:] Micro-emulation of the behaviour of individual agents, creating so called "doppelg\"anger" of each agent.

\item[Theme 2:] Macro-emulation of an entire ABM simulation, using the multivariate time series data.

\item[Theme 3:] Reduction of the complexity to design ANNs, setting the number of input nodes, number of hidden layers, and number of nodes in each hidden layer.

\item[Theme 4:] Reinforcement learning in economic policy design, generating an ANN-policy agent that can be used for policy analysis.
\end{list}

In the end, the emulation of agent-based models using artificial neural networks (ANN-emulation) allows us to perform specific types of analysis that appear quite complicated with large-scale ABMs.\footnote{But not impossible in principle. For example, the global sensitivity analysis of a large-scale ABM such as the Eurace@Unibi Model was already performed using HPC clusters. Also, empirical validation is currently being done for medium-sized ABMs, and given the exponential increase in computing power is expected to yield results in the coming years.} For example, the empirical validation of the ANN-emulator could be an efficient, indirect method to calibrate the parameters of the ABM itself. Also, a global parameter sensitivity analysis could be performed using the ANN-emulator instead of the bulky ABM. And finally, after a careful internal validation procedure to check that the ANN can in fact emulate the ABM to a sufficient degree of accuracy, the ANN-emulator could also be used as a policy analysis tool.

\bigskip
In the following sections we describe each of these themes in more detail.

\subsection{Theme 1: Micro-emulation of the behaviour of economic agents}

This is a local approach, in the sense of modelling the behaviour of each of the rule-based agents by ANNs. A neural network is trained to predict the actions of a particular agent in the model, i.e. the ANN acts as a Doppelganger of that agent. Due to the multitude of instances and agent types, each with their own set of instructions and constraints, and because of the dynamically changing environment of the ABM, such networks can help us model the behaviour of our agents and reduce the complexity of the model at the local level.  The ANNs may need extensive training, but are cheap when they run. In the end, the original agents may be replaced by their Doppelganger, or we may run the hybrid model with both types of agents.

Various decision making problems in standard macroeconomics models are formulated as optimization problems. This theme is dedicated to show that this could be dealt with equally well using machine learning methods. In each example below, we replace a standard optimization problem with a heuristic.

\smallskip
\textit{1. The firm's consumer demand estimation problem.}\\
In the current model, demand is estimated by two methods, one is backward-looking, the other is forward-looking.
In the backward-looking method, the firm only relies on past observations and uses a simple OLS regression on the previous months' sales revenues to estimate the future demand for its product.
It estimates the parameters of a linear, quadratic, or cubic fit for the perceived demand function that the firm beliefs to be facing (for more details, see \citealp[Ch.1, pp 13-14]{Harting:2014}).

In the forward-looking method, the firm uses a market research routine that is commonly used in practice, namely to hold simulated purchase surveys among its consumers (\citealp{NagleHogan:2006}). Once a year, the firm considers a sample of households to present them with a set of products at differing prices. The survey contains questions regarding consumers' preferences and price sensitivities, and asks them how they would react c.q. how they would alter their consumption pattern when the firm would change its price (assuming the prices of all its competitors stay the same). In this way, the firm tries to gauge the overall price sensitivity of consumers, and to estimate its future market share. (see \citealp[Ch. 3, pp. 79-81]{Harting:2014} for more details on the market research method, and references therein).

\smallskip
\textit{The firm's consumer demand estimation problem using artificial neural networks.}\\
The idea of replacing the linear, quadratic, or cubic fitting of the data by a neural network is relatively straightforward. The ANN-firm would try to estimate the local slope of its demand function by way of an ANN, and adjust its arc weights while the simulation is ongoing. Since neural networks are a data-driven approach, there is no need to assume any particular statistical model. Also it is not necessary to rely on linear statistical methods such as OLS, since ANNs are non-linear, non-parameteric time series methods.

\smallskip
\textit{2. The consumers' demand problem.}\\
In the current model, the consumers' decision to buy a product from a specific firm is derived from a discrete choice model using a multinomial logit function. The selection probability to select a firm is an increasing function of the consumer's utility for that firm's product. The utility value is decreasing in price, so a firm with a higher price will have a lower selection probability, but it is bounded away from zero.
By replacing the logit function with a neural network formulation of the same problem, the ANN-consumer can learn how to best achieve its target value. In this way, the consumers' choice problem is redefined as goal-oriented behaviour, rather than as a stochastic model of choice.
 
\smallskip
\textit{3. The banks' internal risk model.}\\
Banks' decision making process involves the problem of setting interest rates for individual loans to the private sector. In order to make a profit, banks should assess their debtors' credit worthiness, and the likelihood that the borrower will not repay the loan in the near future (i.e., the probability that they will default on the credit). This includes an evaluation of the probability of default of the debtor firm, but also the default on individual loans. In order to make such an assessment, the banks either use an internal risk model, or rely on information provided by external rating agencies, or both. Whichever method is being used, they have to assess the various types of risk associated to their assets, including market risk, credit risk, and systemic risk (\citealp{DuffieSingleton:2003}). Market risk refers to changes in the value of the assets on the balance sheet of the bank. Typically, these are fluctuating due to the mark-to-market asset valuation and the volatility of prices on the financial asset markets. Credit risk refers to the risk the bank is facing due to the changing values of assets on the balance sheets of its debtors.

In the current agent-based macroeconomic model (Eurace@Unibi), the bank uses a highly simplified model to determine the probability of default of the firms to which they have outstanding loans, or of new firms that make credit requests. The essential aspect of the model is that the bank's assessment of the firm's probability of default is based on balance-sheet data of the firm, and derived from the firm's equity and financial ratios such as the debt-equity-ratio, an indicator of financial fragility.

Such a "structural-model" approach may or may not be in accordance with the actual behavior of real banks, which would be a matter of empirical study that is beyond the scope of our current research project. But in fact, many alternative models for evaluating credit default risk exist, as illustrated by the rich overview given by \citet{DuffieSingleton:2003}.

One such an alternative approach is the "first-passage model" (\citealp[pp. 53]{DuffieSingleton:2003}), which uses empirical time series collected over a certain time window, to determine the actual default probabilities for a population of firms that have similar risk profiles. Such a time series approach differs substantially from the more theoretical "reduced-form" approaches, but it would fit quite nicely with the neural network approach.

The artificial neural network approach to model the banks' decision making problem will thus provide us with a nonlinear, nonparametric, multivariate time series forecasting method. The bank can be modelled as a goal-oriented entity, that tries to set interest rates based on its forecasted default probabilities, which are derived from time series that are being generated online, i.e. during an ongoing simulation.
In the end, this could yield an agent-based model of the credit market in which the credit risk models proposed in \citet{DuffieSingleton:2003} have been internalized into our agents' behavioural repertoires.

This line of research can distinguish between "offline training" and "online learning". Offline training makes use of time series data being generated by an agent-based model that is "detached" from the agent. One can think of this as an outside-observer-approach, where the agent is not part of the simulation environment, but can observe the actions and outcomes of other agents. This is similar to how children learn how to behave appropriately in a social environment, before they are accepted as full members of that environment.

Online learning, on the other hand, occurs while the agent is itself part of the simulation environment, and is observing the time series being generated online. If multiple agents are simultaneously using online learning in this sense, we can speak of \textbf{co-evolutionary learning} by a population of heterogeneous, artificial neural network agents.

The main aim of this particular research theme is to focus on the appropriate network structure to emulate the multivariate time series data being generated by the target system (in this case, the particular agents to emulate).

The final goals of Theme 1 are to obtain:
(i) a model of firm behaviour replaced by an ANN for the firm's demand estimation routine;
(ii) a model of consumer behaviour replaced by an ANN for the consumers' choice problem;
(ii) a model of bank behaviour replaced by an ANN for the bank's internal risk evaluation and risk-assessment problem.

\subsection{Theme 2: Macro-emulation of an entire ABM simulation model}
Due to the recent breakthrough of Deep Learning techniques for multi-layered networks to model non-linearities, it becomes possible to emulate an entire ABM simulation by an ANN generating time-series. Contrary to the local approach in Theme 1, this is a global approach. A neural network is trained to predict the probabilistic structure of the macro-level variables of the model.

This is useful for robustness and parameter sensitivity analysis, since it allows a much larger exploration of the parameter space. A second advantage is that by training the ANN on many counter-factual scenarios,  it is expected to perform better than an ANN that has been trained just on the empirical, historic data, since this is just a single realization of the empirical data generating process.

In our aim to make the problem of tuning the parameters of an ABM more tractable, we try to emulate the input/output function of the entire ABM by an (ultimately) less complex and more tractable Deep Neural Network. Our starting point is that, in an ABM, a multitude of autonomous agents react to changes in their (economic) environment and, in doing so, alter that very environment. In a Deep Neural Net, a multitude of nodes at different layers can encode different information structures and decision processes, such that the network as a whole can serve specific functions. Bringing the two together, we aim to train a neural network that produces the same output (in terms of time series of macro-economic variables) as the ABM. 

This problem is similar to multi-variate time series forecasting, with the difference that instead of estimating the future values based on the past values, the input to the ABM are the actions of the agents in the model. A recurrent neural network (RNN) is the type of network that is the obvious choice for such a task. A key part of the design is a \emph{feedback} property, namely that the output of the model (the values of the measured macro-economic variables) are fed back to the input.
A second key part is to split the network input layer to represent the `present', and the `past'. This is how the RNN design captures `history': at each step $t$ the network receives inputs at time $t$, but \emph{also} of time $t-1$, and possibly further time lags.

In terms of integrating the decision processes of the individual agents,  a first approximation could be a multi-layered structure in which the nodes of the first input layer are entire ANNs that model each individual agent in the agent-based model. Of course, this is not expected to be any more tractable than the ABM itself. However, it is expected that the ANN will be able to emulate the ABM  \emph{with a much smaller number of agents}, as the multi-layer structure allows the ANN to model increasingly more complex functions of the modelled economy without the need to fully emulate it. Instead of representing the individual agents one by one, the ANN is a representation of the entire ABM at increasing "layers of abstraction". Most importantly, this theme will be informed by other themes, e.g. the modelling in Theme 1 for the individual agent ANNs can inform the initial design of the macro-emulation ANN.

Training data for the macro-emulation ABM will be provided by already collected (synthetic) data from ABM simulations, as well as from new simulations with agents that are themselves ANNs, rather than the current fixed behavioral routines. The big advantage of training the ANN on simulated data, in addition to the abundance of such data, is that such a network will not just \emph{learn} to forecast a specific realization of some ABM emulation, but it will learn the more general underlying data generating mechanism that is common to all such simulations which are seen during the training phase, and, ideally, also for new previously unseen simulations. This provides for an out-of-sample validation stage by using a subset of the synthetic data that was previously unseen by the ANN, and can be used to test the performance of the macro-emulation Deep Neural Network.

The main aim of this particular research theme is to focus on the appropriate network structure to emulate the multivariate time series generated by the macro-ABM as a whole. A second aim is to investigate what is the most appropriate learning/optimization technique for this problem.

The final goal of Theme 2 is to obtain a deep layered ANN that is trained on data generated by an ABM, and that can be usefully applied for empirical validation, and for policy analysis.

\subsection{Theme 3: Reduction of the complexity to design ANNs}

The design and training of deep ANNs is a complex task. To guide the design of the ANN in terms of the number of nodes and hidden layers, and in order to improve the efficiency of the Deep Learning algorithm, the complexity of the ANN must be reduced. 

The problem of training deep neural networks is an unconstrained global minimization problem, i.e. to find the arc weights such that the training error of the ANN is minimized (the training error is the difference between the ANNs performance on the training set and on the test set).

This problem is NP-hard, so the computational costs will increase exponentially with problem size (given by the number of input nodes and the number of hidden layers). Therefore smart heuristics are needed to approximate the global minimum. Many such heuristics have been developed, but most of these assume that the objective function (the loss function or error function) is differentiable in its arguments. Hence the algorithms make use of the gradient and the Hessian of the objective function. Example methods include Gradient Descent (GD), Stochastic Gradient Descent (SGD), Momentum methods and Nesterov's Accelerated Gradient (see \citealp{Sutskever:2013} for an overview, and references therein).

For convex objective functions, to find the global minimum the gradient methods are globally converging, i.e. they will always find the global minimum, but they will just take longer to converge for worse initializations of the parameters. However, for deep and recurrent networks, the initialization does matter since the objective function of such networks cannot be assumed to be convex. Hence, it is important to design good initializations for the algorithms.

The greedy unsupervised pre-training algorithm of \citet{Hinton:2006} and \citet{Hinton:2006b} is a good starting point since it greedily trains the parameters of each layer sequentially. Such greedy layerwise pre-training is then followed by a "fine-tuning" algorithm such as the standard Stochastic Gradient Descent method.

Another method is the Hessian-Free (HF) Optimization (\citealp{Martens:2010, Martens:2012}) that is able to train very deep feed-forward networks even without such a pre-training step. HF is a second-order method and therefore rather slow, but it is very powerful. It is the preferred method of optimization if there is no idea about good initializations of the network.

The most recent innovations in this field, described by \citet[Ch. 7]{Sutskever:2013}, are able to train very deep neural networks (up to 17 hidden layers) by using aggressive Momentum Methods. Such methods use gradient information to update parameters in a direction that is more effective then steepest descent by accumulating speed in directions that consistently reduce the objective function. The most promising method of this type is Nesterov's Accelerated Gradient (NAG) method, which is a first-order optimization algorithm that has better convergence properties than Gradient Descent. It has two parameters: a learning rate $\varepsilon$ and a momentum constant $\mu$, where $(1-\mu)$ can be thought of as the friction of the error surface.
High values of $\mu$ implies the algorithm retains gradient information and leads to fast convergence, while low values imply high friction and a loss of gradient information, leading to slower convergence to the global minimum.

Using NAG with very aggressive momentum values ($\mu$ close to $0.99$) leads to excellent results for problems that previously were deemed unsolvable, such as data sets exhibiting very long time-dependencies (50-200 time steps).

The main aim of this research theme is to focus on Hessian-Free Optimization and Momentum Methods, and possibly adapt these methods to our specific problems.
A second aim is to optimize the choice of network parameters: the number of input nodes, the number of hidden layers, and how many nodes in each hidden layer.

The final goals from Theme 3 are to design good initializations of the network parameters for the Deep Learning algorithms, and to develop insights to inform the optimization methods and the Deep Learning algorithms.

\subsection{Theme 4: Reinforcement learning in economic policy design}
The final theme is to apply a surrogate, or meta-modelling approach to policy decision-making.
A government or central bank agent may be given certain goals (e.g., maintaining a stable price level, or a low unemployment rate, or macrofinancial stability) rather than using hand-crafted rules to serve those goals (such as a Taylor rule for monetary policy). Using reinforcement learning techniques, an agent starts with little knowledge of the world, but given a reward function, the agent learns to perform better over time, during a simulation run. The ABM allows us to evolve successful policies not only by using empirical data, but also by learning from online-generated streaming data. The idea is to have a neural network policy agent (ANN-policy-agent), and this is again a local approach. 

The objective is to develop a model with an endogenous policy-maker, the ANN-policy-agent, who evolves its decision-making routines endogenously. This agent should adapt its policy in response to the behavioural changes of the other agents in the model.

Similar to Theme 1, we can again distinguish between "offline training" and "online learning". 

\subsubsection*{Offline training of the ANN-policy-agent.} We train the ANN-policy-agent using pre-generated, historical data from the original ABM simulation. Its input are (rule-based) policy decisions made during that simulation and time series of economic variables, and we use unsupervised layer-by-layer training. Thus, the ANN-policy-agent learns (unsupervised) the outcome of policy decisions under specific circumstances, and it is possible to re-enforce this training with data from multiple ABM simulations by using a Monte Carlo approach. After this unsupervised pre-training, we perform an additional supervised training phase, in which we reward policy decisions that have desired outcomes. The trained ANN-policy-agent is then used in ABM simulations as the policy-making authority. Depending on the properties and coverage of the training data, this type of ANN-policy-agent is expected to fare well in the test simulations.

\subsubsection*{Online learning by the ANN-policy-agent.} In this setting, the ANN-policy-agent learns from online streaming data. It has to train its weights while taking actual policy actions during a running ABM economy. This situation is bad from an AI point of view, since training normally takes a long time and occurs in isolation of the actual environment. Therefore the analogy to our setup is a ANN-policy-agent that has not been trained before, or has inappropriate weights for the situation. It has to adapt as best it can, similar to a learning child. However, this setting seems more close to what actual real-world policy-makers are facing, especially in times of a changing economic environment. In times of crisis, policy-makers have to adjust quickly to changing circumstances, possibly making choices that appear suboptimal, but satisfying certain target levels.

The main aim of this theme is to focus on which reward functions and what network structures are most appropriate for the ANN-policy-agent.
A second aim is to design the endogenous policy setting behaviour for the ANN-policy-agent: which behavioral heuristics are used, what meta-rules adapt these heuristics, and what are the parameters for the ANN.

The final goal of Theme 4 is to develop a model with a ANN-policy-agent that can set policy endogenously, and is able to adjust its policy response to the behavioral changes of the other agents in the model economy.

\section{Conclusion}
The purpose of this paper is to sketch a line of research in which artificial neural networks (ANNs) are used as computational approximations or as emulators of the nonlinear, multivariate time series dynamics of a pre-existing agent-based model (ABM). In other words, it is a meta-modelling approach using statistical machine learning techniques. There are various advantages to having such an emulator. For instance, it allows for a computationally tractable solution to the issue of parameter sensitivity analysis, robustness analysis, and could also be used for empirical validation and estimation.

The overall goal is to develop new methods and techniques to improve the applicability of macroeconomic ABMs to economic policy analysis.
For the practical implementation of this goal, we need to make advances in two domains:
\begin{itemize}
\item[1.] Deep Learning: developing new machine learning techniques to represent ABMs by ANNs (Themes 1-2).
\item[2.] Complexity Reduction: developing new complexity-reduction techniques to guide the design of ANNs (Theme 3).
\end{itemize}

The work to be done consists of the following broad research themes:
\begin{enumerate}
\item[Theme 1:]
Micro-emulation of the behaviour of agents. A neural network is trained to predict the actions of a particular agent in the model, i.e. the ANN acts as a Doppelganger of that agent.

\item[Theme 2:]
Macro-emulation of an entire ABM simulation. This is a global approach. A neural network is trained to predict the probabilistic structure on the macro-level, of variables in the ABM model, based on the initialization parameters. ANNs have proved to be very successful for multivariate time series forecasting. They are much more flexible than traditional statistical methods since they are nonlinear, nonparametric time series approximation techniques.

\item[Theme 3:]
Reduction of complexity. The design of the structure of the neural network in terms of numbers of input- and output nodes and the number of hidden layers is a complicated problem. In order to improve the efficiency of the Deep Learning algorithm, the complexity of the ANN must therefore be reduced. This can be done by modelling it in terms of a Hamiltonian system, and proceed with describing the time-evolution of the Hamiltonian.

\item[Theme 4:]
Reinforcement learning in policy design. A government or central bank agent may be given certain goals (such as a stable price level, low unemployment rates, or macrofinancial stability), rather than hand-crafted rules. Using reinforcement learning techniques, an agent starts with little knowledge of the world, but given a reward function that models those goals, the agent learns to perform better over time.

This may lead to more flexible policies and more adaptive behaviour on the part of the policy agent, as it allows for a more flexible, discretionary policy setting behavior, rather than using a fixed, rule-based policy. As the policy agent learns how to set policies optimally, it must adapt to the behavioural changes of the other agents, who might change their behaviour in response to the policy. Hence, this policy-feedback-loop addresses in a very natural way the Lucas Critique.
\end{enumerate}

In summary, themes 1 through 4 not only help us to design a strategy how to emulate and estimate agent-based models using artificial neural networks, but it may also contribute to the burgeoning literature on learning in macroeconomics and optimal policy design. Hence, the research programme connects both micro- and macroeconomics, and joins both estimation and emulation in machine learning.

\subsection{Vision and outlook for the future}

When successful, we could apply the new methods to a plethoria of problems. We would have drastically reduced the complexity and computational load of simulating agent-based models, and come up with new methods to model economic agents' behaviour. Furthermore, linking the time series forecasting capabilities of the Deep Learning algorithm to agent-based models also allows us to envision the possibility of docking experiments between different ABMs: the time series output from one ABM can be fed into the Deep Learning algorithm, resulting in an artificial neural network.
This artificial neural network can then be used as an agent inside another, larger-scale ABM. This notion leads to a hierarchical modelling scheme, in which ABMs of ABMs would become feasible. Each agent in the larger ABM can have an internal "mental model" of the world it inhabits, and those mental models can differ to any degree. On the longer term, this approach would allow the inclusion of computational cognitive models into economic agent-based models, allowing the agents to be fully aware of their environment, and to consider the social embedding of their interactions.


\newcommand{\SortNoop}[1]{}

\end{document}